\documentclass[aps,prl,preprintnumbers,showpacs,twocolumn,bibnotes,footinbib,amsmath,amssymb,dvips]{revtex4}

\usepackage{graphicx}
\usepackage{natbib}

\newcommand{\unit}[2]{{#1}~\ensuremath{\mathrm{#2}}}

\begin{document}

\title{New Limits on Coupling of Fundamental Constants to Gravity \\
  Using $^{87}$Sr Optical Lattice Clocks}
\author{S. Blatt}
\email{sebastian.blatt@colorado.edu}
\author{A. D. Ludlow}
\author{G. K. Campbell}
\author{J. W. Thomsen}
\altaffiliation{Permanent address: The Niels Bohr Institute,
  Universitetsparken 5, 2100 Copenhagen, Denmark}
\author{T. Zelevinsky}
\altaffiliation{Current address: Dept. of Physics, Columbia University,
  New York, NY, USA}
\author{M. M. Boyd}
\author{J. Ye}
\affiliation{JILA, National Institute of Standards and Technology and
  University of Colorado, \\
  Department of Physics, University of Colorado, Boulder, CO,
  80309-0440, USA}

\author{X. Baillard}
\author{M. Fouch\'e}
\altaffiliation{Current address: Laboratoire Collisions Agr\'egats
  R\'eactivit\'e, UMR 5589 CNRS, Universit\'e Paul Sabatier Toulouse 3,
  IRSAMC, 31062 Toulouse Cedex 9, France}
\author{R. Le Targat}
\author{A. Brusch}
\altaffiliation{Current address: National
  Institute of Standards and Technology, 325 Broadway, Boulder,
  CO 80305, USA}
\author{P. Lemonde}
\affiliation{LNE-SYRTE, Observatoire de Paris, 61, avenue de l'Observatoire,
  75014, Paris, France}

\author{M. Takamoto}
\author{F.-L. Hong}
\altaffiliation{Permanent address: National Metrology Institute of
  Japan, National Institute of Advanced Industrial Science and
  Technology, Tsukuba, Ibaraki 305-8563, Japan}
\author{H. Katori}
\affiliation{Department of Applied Physics, Graduate School of
  Engineering, The University of Tokyo, Bunkyo-ku, 113-8656 Tokyo,
  Japan}

\author{V. V. Flambaum}
\affiliation{School of Physics, The University of New South Wales,
  Sydney NSW 2052, Australia}

\date{\today}

\begin{abstract}
  The $^1\mathrm{S}_0$-$^3\mathrm{P}_0$ clock transition frequency
  $\nu_\text{Sr}$ in neutral $^{87}$Sr has been measured relative to
  the Cs standard by three independent laboratories in Boulder, Paris,
  and Tokyo over the last three years. The agreement on the $1\times
  10^{-15}$ level makes $\nu_\text{Sr}$ the best agreed-upon optical
  atomic frequency. We combine periodic variations in the $^{87}$Sr
  clock frequency with $^{199}$Hg$^+$ and H-maser data to test Local
  Position Invariance by obtaining the strongest limits to date on
  gravitational-coupling coefficients for the fine-structure constant
  $\alpha$, electron-proton mass ratio $\mu$ and light quark mass.
  Furthermore, after $^{199}$Hg$^+$, $^{171}$Yb$^+$ and H, we add
  $^{87}$Sr as the fourth optical atomic clock species to enhance
  constraints on yearly drifts of $\alpha$ and $\mu$.
\end{abstract}

\maketitle

Frequency is the physical quantity that has been measured with the
highest accuracy. While the second is still defined in terms of the
radio-frequency hyperfine transition of $^{133}$Cs, the higher
precision and lower systematic uncertainty achieved in recent years
with optical frequency standards promises tests of fundamental physics
concepts with increased resolution. For example, some cosmological
models imply that fundamental constants and thus atomic frequencies
had different values in the early universe, suggesting that they might
still be changing. Records of atomic clock frequencies measured
against the Cs standard can be analyzed~\cite{karshenboim05,lea07} to
obtain upper limits on present-day variations of fundamental constants
such as the fine-structure constant $\alpha=e^2/(4\pi\epsilon_0\hbar
c)$ or the electron-proton mass ratio $\mu =
m_e/m_p$~\cite{fischer04,peik04,bize05,peik06,fortier07}. Some
unification theories imply violation of Local Position Invariance by
predicting coupling of these constants to the ambient gravitational
field. Such a dependence could be tested with a deep-space clock
mission~\cite{wolf07}, but would also be observable in the frequency
record of earth-bound clocks as Earth's elliptic orbit takes the clock
through a varying solar gravitational potential~\cite{flambaum07}.
Annual changes in clock frequencies can thus constrain gravitational
coupling of fundamental constants~\cite{bauch02,fortier07,ashby07}.
Good constraints obtained from such analyses require high confidence
in the data and a fast sampling rate. However, a full evaluation of an
atomic clock system takes several days so that high-accuracy frequency
data is naturally sparse.

Three laboratories have measured the doubly forbidden $^{87}$Sr
$^1\mathrm{S}_0$-$^3\mathrm{P}_0$ intercombination line at
$\nu_\text{Sr} = \unit{429\,228\,004\,229\,874}{Hz}$ with high
accuracy over the last three years. These independent laboratories in
Boulder (USA), Paris (France), and Tokyo (Japan) agree at the level of
\unit{1.7}{Hz}~\cite{boyd07,baillard07,campbell07,takamoto06}. The
agreement between Boulder and Paris is $1\times
10^{-15}$~\cite{boyd07,baillard07,campbell07}, approaching the Cs
limit, which speaks for the Sr lattice clock system as a candidate for
future redefinition of the SI second and makes $\nu_\text{Sr}$ the
best agreed-upon optical clock frequency. In this paper, we analyze
the international Sr frequency record for long-term variations and
combine our results with data from other atomic clock species to
obtain the strongest limits to date on coupling of fundamental
constants to gravity. In addition, our data contributes a
high-accuracy measurement of an optical atomic clock species, which
itself has low sensitivity to variation in fundamental constants, to
the search for drifts of fundamental constants, improving confidence
in the null result at the current level of accuracy.

\begin{figure}
  \centering
  \includegraphics[width=\linewidth]{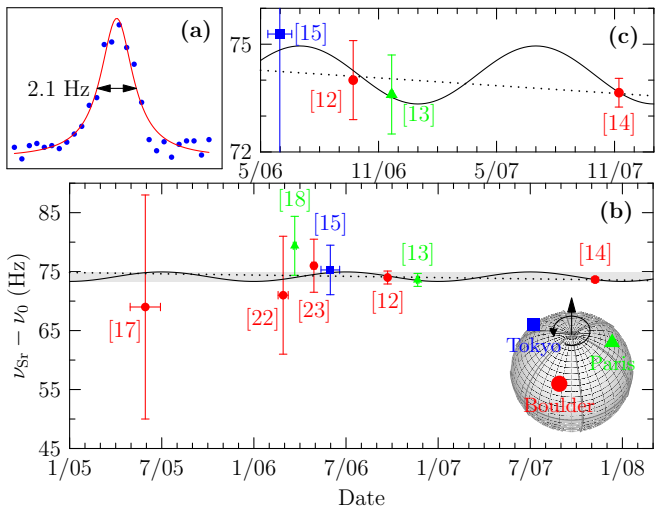}
  \caption{(color online). (a) Spectrum of the $^{87}$Sr
    $^1\mathrm{S}_0$-$^3\mathrm{P}_0$ clock transition with quality
    factor $2\times 10^{14}$. (b) Measurements of clock transition
    from JILA (red circle), SYRTE (green triangle), and U. Tokyo (blue
    square) over the last 3 years. Frequency data is shown relative to
    $\nu_0 = \unit{429\,228\,004\,229\,800}{Hz}$. Weighted linear
    (dotted line) and sinusoidal (solid line) fits determine a yearly
    drift rate and an amplitude of annual variation. (c) Zoom into the
    four most recent measurements, showing agreement within
    $\unit{1.7}{Hz}$ and giving the dominant contribution to both
    drift and annual variation.}
  \label{fig:fit}
\end{figure}

In a strontium lattice clock, neutral fermionic $^{87}$Sr atoms are
trapped at the anti-nodes of a vertical one-dimensional optical
lattice at the Stark-cancellation wavelength, creating an ensemble of
nearly identical quantum absorbers at $\mu$K temperatures. The
$^1\mathrm{S}_0$-$^3\mathrm{P}_0$ clock transition~\cite{katori03} is
interrogated with a highly frequency-stabilized \unit{698}{nm}
spectroscopy laser in the resolved sideband limit and the Lamb-Dicke
regime~\cite{ludlow06,letargat06,takamoto06,boyd07,baillard07,campbell07}.
Using individual magnetic sublevels, spectra with quality factors of
$>2\times 10^{14}$ have been recovered~\cite{boyd06} as shown in
Fig.~\ref{fig:fit}(a). This high-resolution spectroscopy afforded by
the optical lattice allows measurement of the clock frequency with
high accuracy and evaluation of systematic uncertainties at one part
in $10^{16}$, limited by blackbody and residual density
effects~\cite{ludlow08}. Spectroscopic information from the atomic
sample is used to steer the laser to match the clock transition
frequency, which is then measured relative to the Cs standard using an
octave-spanning optical frequency comb~\cite{foreman07}.

In combination with data from other optical atomic clock species,
variations in the measured Sr clock frequency can constrain variation
of fundamental constants. It is necessary to analyze a diverse
selection of atomic species to rule out species-dependent systematic
effects and test the broad predictions of the underlying relativistic
theory. We will introduce the formalism required to constrain the
coupling to gravity by first analyzing the global frequency record for
linear drifts in $\alpha$ and $\mu$.

Figure~\ref{fig:fit}(b) displays Sr clock frequency measurements since
2005. The frequency uncertainties are based on values from
references~\cite{ludlow06,eftf06,letargat06,takamoto06,icap06,boyd07,baillard07,campbell07}.
The date error bar indicates the time interval over which each
measurement took place. A weighted linear fit (dotted line) results in
a frequency drift of $(-1.0 \pm 1.8)\times 10^{-15}/\text{yr}$, mostly
determined by the difference between the last three high-accuracy
measurements~\cite{boyd07,baillard07,campbell07}. This yearly drift
can be related to a drift of fundamental constants via relativistic
sensitivity constants $K_\text{rel}$. Values for various clock
transitions of interest have been calculated in
references~\cite{angstmann07,flambaum06} and the fractional frequency
variation of an optical transition can be written as
\begin{equation}
  \label{eq:2}
  \frac{\delta\nu_\text{opt}}{\nu_\text{opt}} = \
  K_\text{rel}^\text{opt}\frac{\delta\alpha}{\alpha}.
\end{equation}

\begin{figure}
  \centering
  \includegraphics[width=\linewidth]{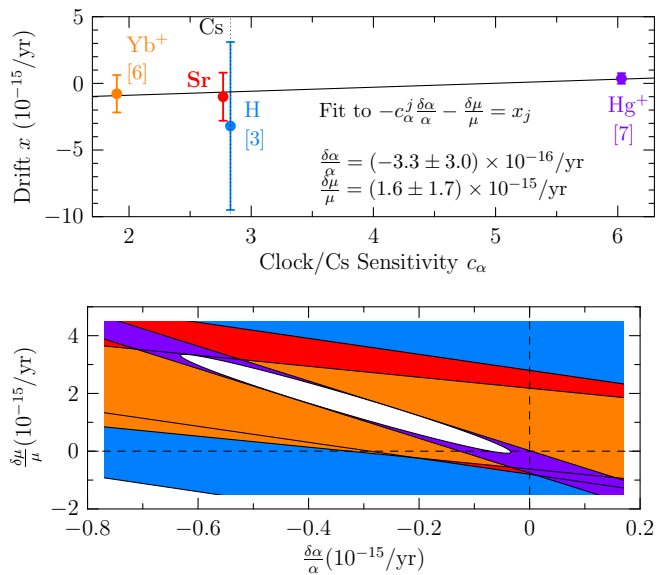}
  \caption{(color online). The upper panel shows fractional frequency
    drifts for $^{171}$Yb$^+$ (orange), $^{87}$Sr (red), H 1S-2S
    (blue) and $^{199}$Hg$^+$ (purple) versus their sensitivity to
    $\alpha$-variation relative to Cs. Sensitivity due to Cs is
    indicated as a dotted vertical line. A linear fit (solid line)
    determines yearly drift rates $\delta\alpha/\alpha$ and
    $\delta\mu/\mu$. The drift rate constraints from each species are
    shown in the lower panel as respectively colored bars. The fit
    determines a confidence ellipse
    (white)~\cite{fischer04,peik06,fortier07} with projections equal
    to the parameters' 1-$\sigma$ uncertainties.}
  \label{fig:drift}
\end{figure}

The Cs standard operates on a hyperfine transition, which is also
sensitive to variations in $\mu$. For a hyperfine transition, the
above equation is modified to
\begin{equation}
  \label{eq:3}
  \frac{\delta\nu_\text{hfs}}{\nu_\text{hfs}} =
  (K_\text{rel}^\text{hfs} + 2)
  \frac{\delta\alpha}{\alpha} + \frac{\delta\mu}{\mu}.
\end{equation}
Here, the change in $\mu$ arises from variations in the nuclear
magnetic moment of the Cs atom~\cite{flambaum06}. The following drift
analysis will focus on optical clocks measured against Cs, since
inclusion of hyperfine clock data from Rb/Cs~\cite{marion03,bize05}
does not change the results significantly.

The overall fractional frequency variation $x_\text{\it j}$ of an
optical clock species $j$ compared to Cs can be related to variation
of $\alpha$ and $\mu$ as
\begin{equation}
  \label{eq:4}
  \begin{split}
  x_\text{\it j} \equiv \frac{\delta(\nu_\text{\it j}/\nu_\text{Cs})}
  {\nu_\text{\it j}/\nu_\text{Cs}} &= \left(K_\text{rel}^{\it j}-
    K_\text{rel}^\text{Cs}-2\right)\frac{\delta\alpha}{\alpha} -
  \frac{\delta\mu}{\mu} \\
  &\equiv -c_\alpha^{\,j}\frac{\delta\alpha}{\alpha} - \frac{\delta\mu}{\mu}.
  \end{split}
\end{equation}
For $^{87}$Sr in particular, $-c_\alpha^{\,\text{Sr}} = 0.06 - 0.83 -
2 = -2.77$~\cite{angstmann07}. The $^{87}$Sr sensitivity is about 50
times lower than that of Cs, so that our measurements are a clean test
of the Cs frequency variation. This allows Sr clocks to serve a
similar role as H in removing the Cs contribution from other optical
clock experiments or to act as an anchor in direct optical
comparisons~\cite{fischer04}.

Other optical clock species with different sensitivity constants have
also been analyzed for frequency drifts. Each species becomes
susceptible to variations in both $\alpha$ and $\mu$ by referencing to
Cs. Figure~\ref{fig:drift} shows current optical frequency drift rates from
Sr, Hg$^+$~\cite{fortier07}, Yb$^+$~\cite{peik06}, and H~\cite{fischer04}. Linear
regression~\cite{zimmermann05} limits drift rates to
\begin{equation}
\begin{split}
  \delta\alpha/\alpha &= (-3.3 \pm 3.0)\times 10^{-16}/\text{yr}\\
  \delta\mu/\mu &= (1.6 \pm 1.7)\times 10^{-15}/\text{yr},
\end{split}
\end{equation}
decreasing the H-Yb$^+$-Hg$^+$~\cite{fischer04,peik06,fortier07}
errorbars~\cite{kolachevsky07} by $\sim$15\% and confirming the null
result at the current level of accuracy by adding high-accuracy data
from a very insensitive species such as Sr to Fig.~\ref{fig:drift}. We
note that another limit on $\delta\alpha/\alpha$ independent of other
fundamental constants (using microwave transitions in atomic Dy) has
recently been reported as $\unit{(-2.7\pm
  2.6)}{10^{-15}/\text{yr}}$~\cite{cingoz07}.

\begin{figure}
  \centering
  \includegraphics[width=\linewidth]{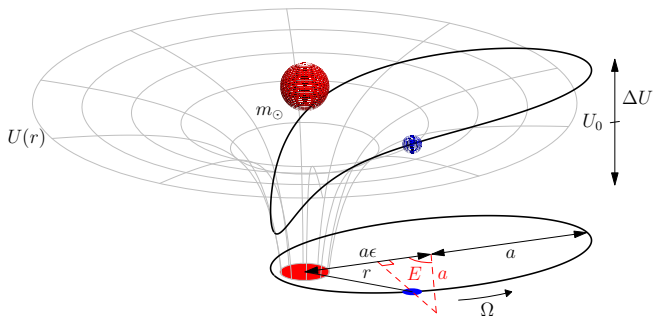}
  \caption{(color online). Earth (blue) orbiting around Sun (red, mass
    $m_\odot$) in gravitational potential $U$ on an orbit with
    semi-major axis $a$, eccentricity $\epsilon$ (exaggerated to show
    geometry) and angular velocity $\Omega$. Earth is shown at radial
    distance $r$ from the Sun. The eccentric anomaly $E$ is the angle
    between the major axis and the orthogonal projection of Earth's
    position onto a circle with radius $a$.}
  \label{fig:orbit}
\end{figure}

We will now generalize the formalism used for the analysis of linear
drifts to constrain coupling to the gravitational potential $U$ and
search for periodic variations in the global frequency record. The
dominant contribution to changes in the ambient gravitational
potential is due to the ellipticity of Earth's orbit around the Sun.
Suppose that the variation of a fundamental constant $\eta$ is related
to the change in gravitational potential via a dimensionless coupling
constant $k_\eta$~\cite{flambaum07}:
\begin{equation}
  \label{eq:6}
  \frac{\delta\eta}{\eta} \equiv k_\eta \frac{\Delta U(t)}{c^2},
\end{equation}
where $\Delta U(t) = U(t) - U_0$ is the variation in the gravitational
potential versus the mean solar potential on earth $U_0$, and $c$ is
the speed of light.

\begin{figure}
  \centering
  \includegraphics[width=\linewidth]{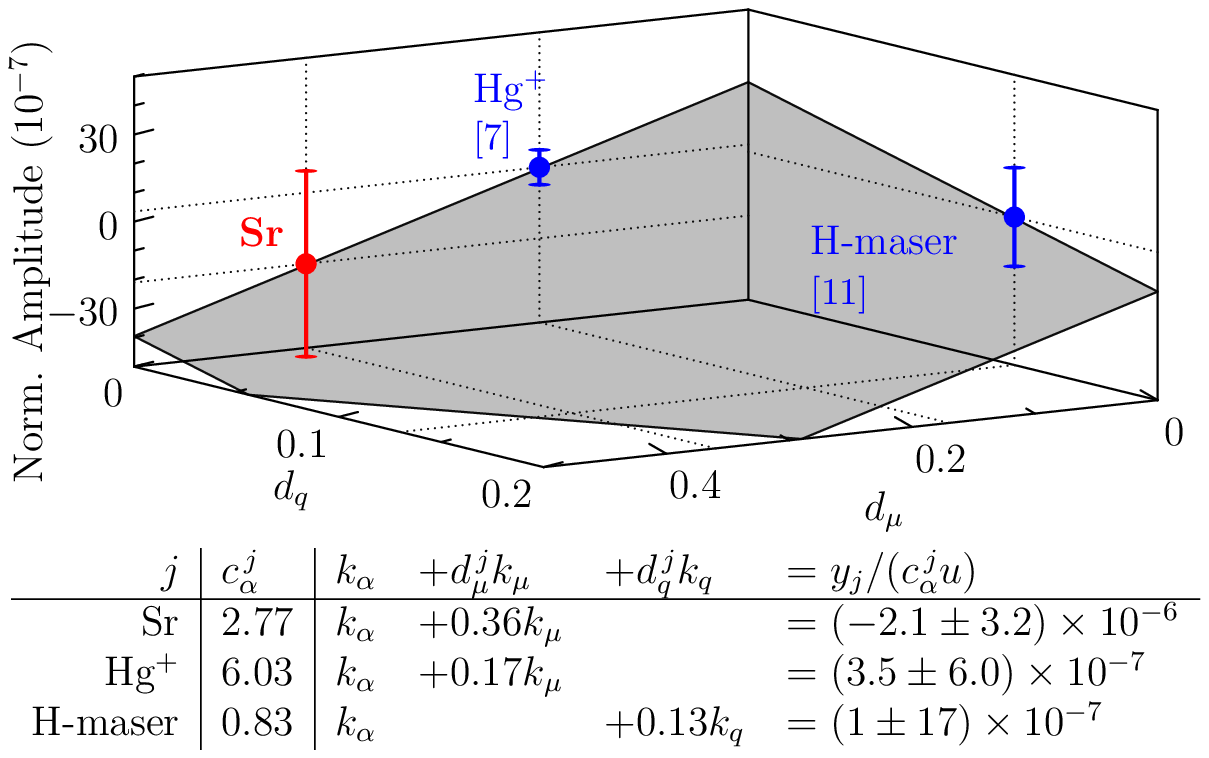}
  \caption{(color online). A fit to linear constraints on gravitational
    coupling constants $k_\alpha$, $k_\mu$ and $k_q$ from three species
    determines a plane. Its value at $d_\mu$=0, $d_q$=0 is $k_\alpha$;
    its gradient along the $d_\mu$ ($d_q$) axis is $k_\mu$ ($k_q$). The
    table shows sensitivity constants and constraints for $^{87}$Sr,
    $^{199}$Hg$^+$ and the H-maser.}
  \label{fig:grav}
\end{figure}

The variation in solar gravitational potential can then be estimated
from Earth's equations of motion (see Fig.~\ref{fig:orbit}). Since
Earth's orbit is nearly circular, we expand the solar gravitational
potential $U(t) = -G m_\odot/r(t)$, with gravitational constant $G$,
Sun mass $m_\odot$, and radial distance Earth--Sun $r(t)$, in the
orbit's ellipticity $\epsilon \simeq 0.0167$. Kepler's
equation~\cite{smart53} relates the eccentric anomaly $E \equiv
\arccos[(1-r/a)/\epsilon]$ (with semi-major axis $a \simeq
\unit{1}{au}$) to the orbit's elapsed phase since perihelion:
\begin{equation}
  \label{eq:9}
  \Omega t = E - \epsilon\sin{E},
\end{equation}
where $\Omega \simeq \sqrt{G m_\odot /a^3} \simeq \unit{2\times
  10^{-7}}{s^{-1}}$ is Earth's angular velocity from Kepler's third
law. Kepler's equation has a solution given by a power series in the
ellipticity as $E=\Omega t + \mathcal{O}(\epsilon)$, which can be used
to expand $1/r$ and thus $\Delta U$ to first order in $\epsilon$:
\begin{equation}
\Delta U(t) = -\frac{G m_\odot}{a}\epsilon \cos{\Omega t},
\end{equation}
with a dimensionless peak-to-peak amplitude $u\equiv 2 G m_\odot / (a
c^2) \simeq 3.3 \times 10^{-10}$. Thus, the $^{87}$Sr fractional
frequency variation due to gravitational coupling is
\begin{equation}
  \label{eq:13}
  x_\text{Sr}(t) = \left[2.77 k_\alpha + k_\mu\right]
  \frac{G m_\odot}{a c^2} \epsilon \cos{\Omega t},
\end{equation}
with amplitude containing $k_\alpha$ and $k_\mu$ as the only free
parameters. Fitting Eqn.~\ref{eq:13} to the combined Sr frequency
record in Fig.~\ref{fig:fit}(b) gives an annual variation with
amplitude $y_\text{Sr} = (-1.9 \pm 3.0)\times 10^{-15}$, which
constrains $2.77 k_\alpha + k_\mu$ by division through $u$.

Other atomic clock species that have been tested for gravitational
coupling are $^{199}$Hg$^+$~\cite{fortier07} and the
H-maser~\cite{ashby07}. H-masers are also sensitive to variations in
the light quark mass~\cite{flambaum06}, adding a third coupling
constant $k_q$. Although the maser operates on a hyperfine transition,
the H atom is well understood, permitting the use of H-maser data with
optical clocks to constrain $k_q$. Using sensitivity coefficients from
references~\cite{angstmann07,flambaum06}, each atomic clock species
$j$ contributes a constraint of the general form~\cite{flambaum07}
\begin{equation}
  \label{eq:1}
  c_\alpha^{\,j} k_\alpha + c_\mu^{\,j} k_\mu + c_q^{\,j} k_q =
  y_\text{\it j}/u.
\end{equation}
Division by $c_\alpha^{\,j}$ gives this equation the form of a linear
function in two variables $d_\mu^{\,j} \equiv
c_\mu^{\,j}/c_\alpha^{\,j}$ and $d_q^{\,j} \equiv
c_q^{\,j}/c_\alpha^{\,j}$.

In Fig.~\ref{fig:grav}, each species' constraint is interpreted as a
measurement of this linear function in the numerical
coefficients~\footnote{The constraint for Hg$^+$ is corrected for a
  sign error in applying Eqn.~2 of reference~\cite{fortier07} in the
  subsequent paragraph. The sign of the constraint for the H-maser
  derives from the averaged fit in Fig.~3 of
  reference~\cite{ashby07}.}. A linear fit gives:
\begin{equation}
\begin{split}
  k_\alpha &= (2.5 \pm 3.1)\times 10^{-6}\\
  k_\mu &= (-1.3 \pm 1.7)\times 10^{-5} \\
  k_q &= (-1.9 \pm 2.7)\times 10^{-5}.
\end{split}
\end{equation}
Due to the orthogonal dependence on $k_q$, the maser data only pivots
the plane in Fig.~\ref{fig:grav} around the Hg$^+$--Sr line, but its
value and error bar influence neither the value nor the error bar of
$k_\alpha$ and $k_\mu$. The values agree well with zero and we
conclude that there is no coupling of $\alpha$, $\mu$ and the light
quark mass to the gravitational potential at the current level of
accuracy. We note that the coupling constant $k_\alpha$ has recently
been measured independently in atomic Dy, resulting in $k_\alpha =
(-8.7\pm 6.6)\times 10^{-6}$~\cite{ferrell07}, limited by systematic
effects. While optical clocks are not as sensitive to variations in
constants as Dy, systematic effects have been characterized at much
higher levels~\cite{ludlow08}.

The unprecedented level of agreement between three international labs
on an optical clock frequency allowed precise analysis of the Sr clock
data for long-term frequency variations. We have presented the best
limits to date on coupling of fundamental constants to the
gravitational potential. In addition, by adding a high-accuracy
measurement of a low-sensitivity species to the analysis of drifts of
fundamental constants, we have increased confidence in the zero drift
result for the modern epoch.

The Boulder group thanks T. Ido, S. Foreman, M. Martin and M. de
Miranda as well as T. Parker, S. Diddams, S. Jefferts and T. Heavner
of NIST for technical contributions and discussions. The Paris
group acknowledges contributions by P. G. Westergaard, A.
Lecallier, the fountain group at LNE-SYRTE, and by G. Grosche, B.
Lipphardt and H. Schnatz of PTB. The Tokyo group thanks Y. Fujii and
M. Imae of NMIJ/AIST for GPS time transfer. We thank S. N. Lea for
helpful discussions.

Work at JILA is supported by ONR, NIST, NSF and NRC. SYRTE is Unit\'e
Associ\'ee au CNRS (UMR 8630) and a member of IFRAF. Work at LNE-SYRTE
is supported by CNES, ESA and DGA. Work at U. Tokyo is supported by
SCOPE and CREST.

\end{document}